# Optical investigation of electronic states of $Mn^{4+}$ ions in *p*-type GaN

B. Han and B. W. Wessels[a]

*Department of Materials Science and Engineering and Materials Research Center,*

*Northwestern University, Evanston, Illinois 60208, USA*

M. P. Ulmer

*Department of Physics and Astronomy, Northwestern University, Evanston, Illinois 60208*

The electronic states of manganese in *p*-type GaN are investigated using photoluminescence (PL) and photoluminescence excitation (PLE) spectroscopies. A series of sharp PL lines at 1.0 eV is observed in codoped GaN and attributed to the intra *d*-shell transition $^4T_2(F)$-$^4T_1(F)$ of $Mn^{4+}$ ions. PLE spectrum of the $Mn^{4+}$ [$^4T_2(F)$-$^4T_1(F)$] luminescence reveals intra center excitation processes via the excited states of $Mn^{4+}$ ions. PLE peaks observed at 1.79 and 2.33 eV are attributed to the intra-d-shell $^4T_1(P)$-$^4T_1(F)$ and $^4A_2(F)$-$^4T_1(F)$ transitions of $Mn^{4+}$, respectively. In addition to the intra shell excitation processes, a broad PLE band involving charge-transfer transition of the $Mn^{4+/3+}$ deep level is observed, which is well described by the Lucovsky model. As determined from the onset of this PLE band, the position of the $Mn^{4+/3+}$ deep level is 1.11 eV above the valence band maximum, which is consistent with prior theory using *ab initio* calculations. Our work indicates 4+ is the predominant oxidation state of Mn ions in *p*-type GaN:Mn when the Fermi energy is lower than 1.11 eV above the valence band maximum.

[a]Electronic Mail: b-wessels@northwestern.edu

Transition metals in III-V semiconductors have been previously extensively studied because of their scientific importance and technological applications.[1,2] Recently there has been renewed interest due to the discovery of ferromagnetism in semiconductors alloyed with magnetic elements.[3,4] One system that has received considerable attention is manganese in GaN.[5-7] This system has been reported to be a ferromagnetic semiconductor with a high Curie temperature due to a large *p-d* hybridization and small spin-orbit coupling.[4-6,8] The ferromagnetic properties of GaN:Mn depend strongly on the electronic configuration of the Mn ions,[9,10] which is determined by the oxidation state of Mn and the local crystal field.[1,2] The 2+ ($d^5$) and 3+ ($d^4$) oxidation states are typically observed for substitutional manganese in III-V semiconductors. Both the $d^4$ and $d^5$ configurations have been observed by electron paramagnetic resonance and optical studies.[11-14] The $Mn^{3+/2+}$ acceptor level was determined to be 1.4-1.8 eV above the valence band maximum.[11,12]

Recent *ab initio* calculations indicated that besides $Mn^{2+}$ and $Mn^{3+}$, the oxidation state of $Mn^{4+}$ could also be stabilized in wurtzite GaN (with a tetrahedral local crystal field) in that the $Mn^{4+/3+}$ transition level is 1.1 eV above valence band maximum.[15] However, the position of the $Mn^{4+/3+}$ level in GaN has not been experimentally determined; the electronic states of $Mn^{4+}$ haven't been established as well. It should be noted that the $Mn^{4+}$ oxidation state has not been previously observed in a tetrahedral crystal field in that the $Mn^{4+/3+}$ transition level is predicted to be resonant with the valence band in most semiconductors.[1,2]

In this work the optical properties of GaN:Mn-Mg are investigated by photoluminescence (PL) and photoluminescence excitation (PLE) spectroscopies to determine the $Mn^{4+/3+}$ energy level position, as well as the excited state energy levels of the $Mn^{4+}$ ions. A series of sharp PL lines at 1.0 eV is observed in *p*-type GaN:Mn heavily codoped with the acceptor Mg. Their peak energies remain invariant with increasing temperature up to 200 K,[7] which is characteristic of intra d-shell transitions.[1,2] From the transient decay properties of the 1.0 eV PL, they are attributed to the spin allowed intra shell $^4T_2(F)$-$^4T_1(F)$ transitions of $Mn^{4+}$ ions due to their short lifetimes of 75±10 μsec.[7] In this work by examining the excitation behavior of the $^4T_2(F)$-$^4T_1(F)$ PL transition using PLE spectroscopy, the position of the $Mn^{4+/3+}$ deep level is determined and compared to *ab initio* calculations.[15] The energy levels of the $^4T_2(F)$, $^4T_1(P)$, and $^4A_2(F)$ excited states are also measured.

Semi-insulating, *p*-type GaN:Mn epilayers codoped with Mg were grown by atmospheric pressure metal-organic vapor phase epitaxy as previously described.[11,16,17] For PLE measurements a 250 W quartz-halogen lamp was used as the excitation source, spectrally dispersed by a 0.75m SPEX grating monochromator. The luminescence was detected by a cooled Ge detector. A Zeiss MM12 double prism monochromator is employed as a band-pass filter to define the detection window. The PLE spectra were normalized to account for the spectral responsivity of the lamp and the SPEX monochromator. For PL spectroscopy the excitation source was either the 325 nm line of a He-Cd laser, or the 973.8 nm line of a semiconductor laser. The sample temperature was varied between 20 and 300 K using a closed-cycle helium cryostat.

Figure 1 shows the 20K PL spectra of codoped GaN pumped by (a) below gap and (b) above gap laser excitation. A series of sharp peaks at 1.0 eV dominates both spectra with the full width at half maximum of 3-8 meV. Although positions of the major sharp PL peaks are independent of excitation wavelength (indicated by the dotted lines in Fig. 1), a strong dependence of their relative intensities on the excitation wavelength is observed, indicating these PL transitions are related to different $Mn^{4+}$ centers formed in codoped GaN.[7] A series of lower intensity PL peaks are also observed on the low energy side of the PL spectra, which are attributed to phonon replicas of the major peaks. In Fig. 1, the major peaks and their corresponding phonon replicas are linked by the arrows. The phonon energy is determined to be 64±1 meV.

To determine the excitation mechanisms for the 1.0 eV PL peaks, as well as the electronic states of the $Mn^{4+}$ ions in codoped GaN, the PLE spectra of the $Mn^{4+}$ [$^4T_2$(F)-$^4T_1$(F)] luminescence were measured at low temperature. When measuring the PLE spectra, the $Mn^{4+}$ luminescence was detected in a 70-meV window around the PL peaks at 1.03 eV. Figure 2(a) shows the 20 K PLE spectrum of the 1.03 eV peak. A sharp increase of the PLE response with decreasing photon energy is observed at the low energy side ($h\nu < 1.15$ eV), which results when the energy of the dispersed excitation photons is tuned close to the detection window around 1.03 eV. The shape of the PLE spectrum indicates the presence of a broad excitation band with the threshold energy ~1.2 eV. Due to its large spectral bandwidth, this band is attributed to a PLE transition between a deep impurity level and the conduction or valence band (i.e. the photoionization of a deep level). Assuming the conduction/valence band to have a parabolic shape, the spectral dependence of the photoionization cross section can be determined using a Lucovsky model whereby $\sigma_\nu \sim (h\nu - E_i)^{3/2}/(h\nu)^3$, where $h\nu$ is the photon energy and $E_i$ is the ionization energy of the deep level.[18] By fitting the PLE band using the Lucovsky model [shown as the dashed line in Fig. 2(a)], the ionization energy of the deep level is determined to be 1.11 eV. It should be noted that the observed ionization energy is in good agreement with *ab initio* calculations for Mn in GaN, where the $Mn^{4+/3+}$ transition level is calculated to be 1.1 eV above the VBM.[15] Consequently, the PLE band with the $E_i$ of 1.11 eV is attributed to the photoionization of holes from the $Mn^{3+}$ ions to the valence band.

The PLE processes of this $Mn^{4+/3+}$ broad band are summarized in Eq. (1):

$$Mn^{4+}[^4T_1(F)] + h\nu \rightarrow Mn^{3+} + h^+ \rightarrow Mn^{4+}[^4T_2(F)] \rightarrow Mn^{4+}[^4T_1(F)] + h\nu_0 \quad (1)$$

Holes are directly excited from the $Mn^{4+}[^4T_1(F)]$ ground state to the valence band when the photon energy is larger than 1.11 eV, forming $Mn^{3+}$ ions. The recombination of holes with the $Mn^{3+}$ centers subsequently excites the 3d shell electrons to $Mn^{4+}[^4T_2(F)]$ states, followed by the $^4T_2(F)$-$^4T_1(F)$ PL transitions of $Mn^{4+}$ ions at 1.0 eV (Fig. 1). A similar excitation process involving charge transfers has been described for intra d-shell transition of $Fe^{3+}$ ions in GaN.[19]

Three other bands at 1.79, 2.33 and 3.3 eV are distinctly observed upon subtracting the broad band from the PLE spectrum, as shown in Fig. 2(b). The intensities of the two peaks at 1.79 and 2.33 eV decrease with increasing photon energy after reaching the maximum, which is characteristic of an intra center transition. Consequently they are attributed to the intra *d*-shell excitations of $Mn^{4+}$ ions. According to the Tanabe-Sugano diagram of a $d^3$ ion ($Mn^{4+}$) in the tetrahedral crystal field (GaN), the 1.79 and 2.33 eV PLE peaks are attributed to spin allowed intra *d*-shell transitions from the ground $^4T_1(F)$ state to the $^4T_1(P)$ and $^4A_2(F)$ excited states, respectively.[2] The intra *d*-shell excitation processes can be explained as follows: when the pump energy is tuned resonantly with one of its excited states, the $Mn^{4+}$ ion is excited from the ground state to the corresponding excited state. The excited $Mn^{4+}$ ion subsequently relaxes non-radiatively to the first excited $^4T_2(F)$ state before the $^4T_2(F)$-$^4T_1(F)$ PL transition at 1.0 eV.

From the PLE spectrum of the $Mn^{4+}$ [$^4T_2(F)$-$^4T_1(F)$] luminescence, a comprehensive electronic diagram of the $Mn^{4+}$ ions in hexagonal GaN can be constructed and is shown in Fig. 3. Four crystal-field states of $Mn^{4+}$ ions are identified, whose energy positions give a wealth of information for detailed crystal-field calculations that are beyond the scope of this paper.

It is of interest to note that the energy separation between the $^4T_2(F)$ and $^4T_1(F)$ states of $Mn^{4+}$ in GaN is 1.0 eV, approximately twice as large as the energy separation of $V^{2+}$ (also $3d^3$ configuration) in the tetrahedral crystal field of II-VI compounds;[20,21] i.e. the crystal field splitting of the $^4F$ multiplet of $Mn^{4+}$ in GaN is twice of the splitting for $V^{2+}$ in II-VI compounds.[1,2] The observed stronger crystal field is presumably a combined result of i) the small lattice constant of GaN, ii) the 4+ charge state of manganese, and iii) the high electron affinity of nitrogen.[19] A large crystal-field splitting has been observed for other transition metals in GaN.[19,22]

In addition to the three PLE peaks discussed above, another band with the threshold energy of 3.3 eV is also observed. The intensity of this peak increases monotonically with increasing photon energy, which is characteristic of an excitation process between a discrete level and the conduction or valence band. From the threshold energy of 3.3 eV, this band is attributed to photoionization of electrons from negatively charged $Mg_{Ga}^-$ acceptors to the conduction band in that the $Mg_{Ga}^{0/-}$ (or $Mg^{3+/2+}$) level is 0.2 eV above the valence band maximum[23] and the low temperature bandgap energy of GaN is ~3.5 eV. This attribution is consistent with the presence of high concentration (~ $10^{19}$ cm$^{-3}$) $Mg^{2+}$ ions in our films.[23,24] The PLE processes of 3.3 eV band can be described as follows: electrons are excited from negatively charged $Mg_{Ga}^-$ (i.e. $Mg^{2+}$) ions to the conduction band for a photon energy greater than 3.3 eV. The free electron is subsequently trapped at the $Mn^{4+}$ related center. The electron occupied trap then captures a hole and a bound exiton is formed at the $Mn^{4+}$ center. Recombination of this bound exiton excites the 3d shell electrons of $Mn^{4+}$ to the $^4T_2(F)$ state through an Auger type process.[25] The observed 1.0 eV peak is a result of the $^4T_2(F)$-$^4T_1(F)$ radiative transition of $Mn^{4+}$. Alternatively the 3.3 eV PLE band can be explained by a Dexter energy transfer process.[26,27,28]

Finally it should be noted that since the $Mn^{4+}$ ions have a $d^3$ configuration the magnetic properties of the *p*-type codoped GaN:Mn would be expected to be different than material which has predominantly $Mn^{2+}$ ($d^5$) or $Mn^{3+}$ ($d^4$) ions.[9-12]

In summary, the optical properties of Mn ions in the 4+ oxidation state in tetrahedral crystal field (of *p*-type GaN:Mn-Mg) were measured. The electronic states of $Mn^{4+}$ ions in codoped GaN:Mn-Mg are determined by PLE spectroscopy of the $Mn^{4+}$ [$^4T_2(F)$-$^4T_1(F)$] luminescence. The $Mn^{4+/3+}$ level is 1.11 eV above the VBM, which is in excellent agreement with the *ab initio* calculations. Two intra *d*-shell transitions at 1.79 and 2.33 eV are observed in the PLE spectrum and attributed to the photon excitations of the $Mn^{4+}$ ions from the ground state $^4T_1(F)$ to different excited states $^4T_1(P)$ and $^4A_2(F)$, respectively. The $Mn^{4+}$ [$^4T_2(F)$-$^4T_1(F)$] luminescence can also be excited by photon ionization of nearby $Mg^{2+}$ ions through an Auger process.

This work is supported by NASA Grant No. NAG5-1147, NSF SPINS program under Grant No. ECS-0224210, and ONR Grant No. N00014-01-0012.

[28] In the Dexter model, electrons are excited from negatively charged $Mg_{Ga}^-$ (i.e. $Mg^{2+}$) ions to the conduction band when the photon energy is greater than 3.3 eV, resulting in the formation of $Mg^{3+}$ ions. $Mg^{2+}$ bound exitons are subsequently formed when the electrons excited to the conduction band are trapped by the $Mg^{3+}$ ions. The energy released from recombination of the bound exitons is then transferred by a Dexter non-radiative mechanism to nearby $Mn^{4+}$ ions [D. L. Dexter, J. Chem. Phys. **21**, 836 (1953)], inducing an excitation of $Mn^{4+}$ ions from the ground $^4T_1(F)$ state to one of the excited states. The excited $Mn^{4+}$ ions subsequently relax non-radiatively to the first excited state $^4T_2(F)$, followed by the 1eV PL emission. Similar excitation process has been observed for $Er^{3+}$ in GaN and Mg codoped GaN [S. Kim, S. J. Rhee, X. Li, J. J. Coleman, and S. G. Bishop, Appl. Phys. Lett. **76**, 2403 (2000)].

**Figure captions:**

FIG. 1. Comparison of the intra d-shell transition of $Mn^{4+}$ ions in the PL spectra excited by (a) below gap (973.8 nm semiconductor) and (b) above gap (325 nm He-Cd) laser lines. The major PL peaks and their phonon replicas are linked by arrows.

FIG. 2. (a) The calibrated PLE spectrum of codoped GaN:Mn-Mg at 20 K; (b) the 20 K PLE spectrum after subtracting the Lucovsky fit.

FIG. 3. Energy levels of $Mn^{4+}$ ions in wurtzite GaN. The energies are given for a crystal temperature of 20 K.

Figure 1

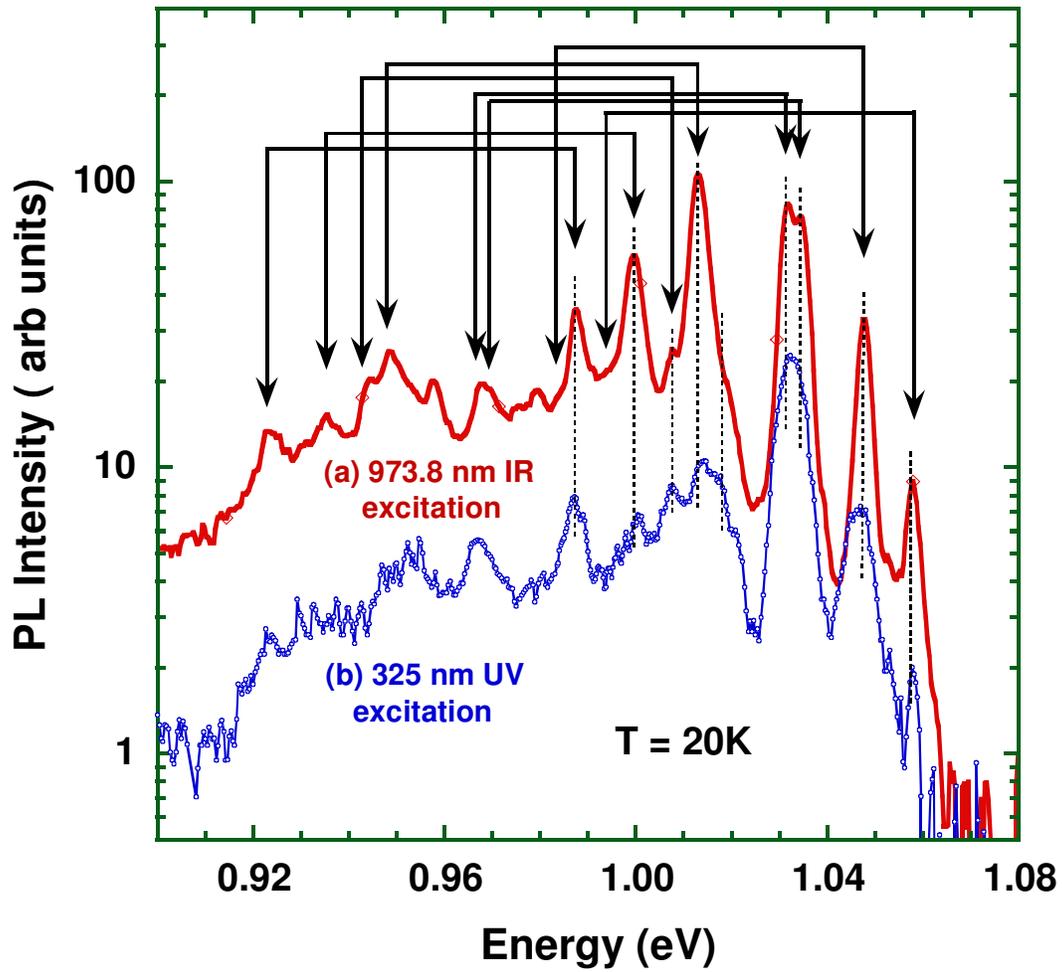

Figure 2

(a)

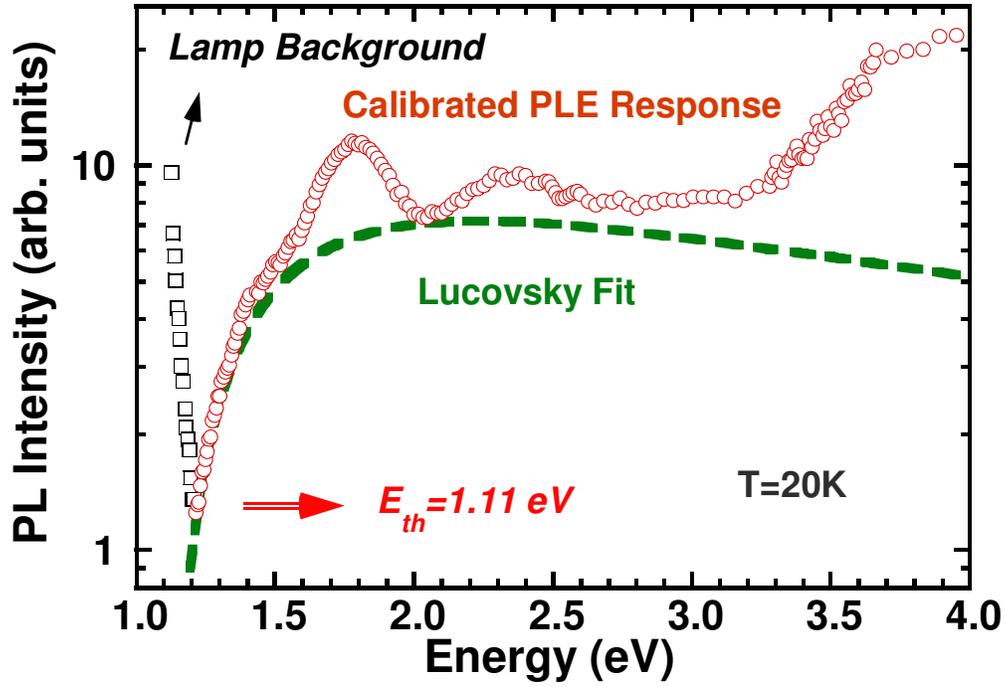

(b)

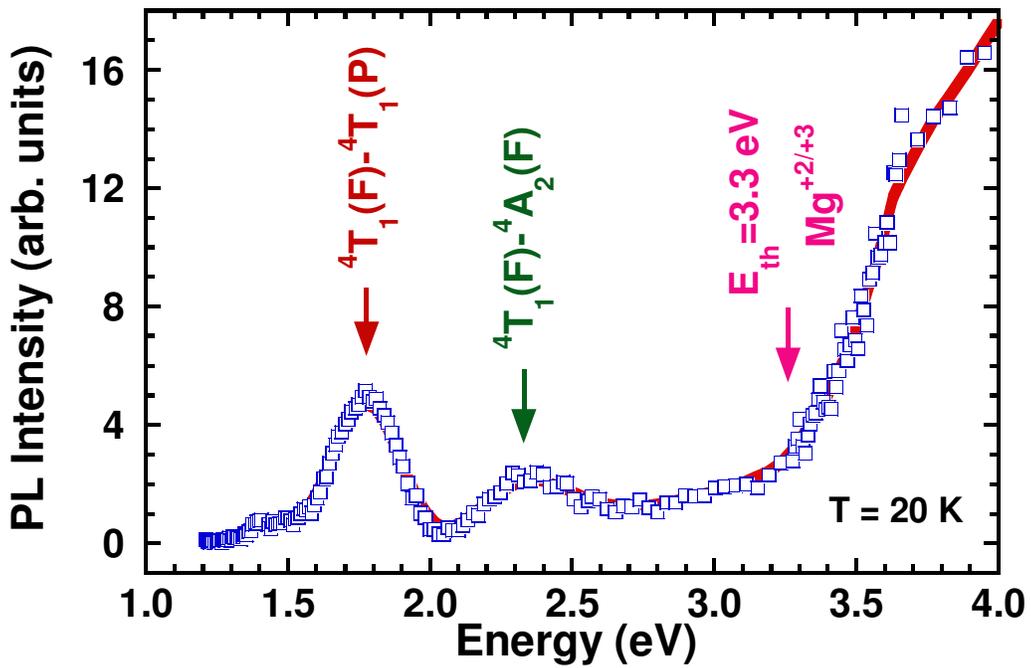

Figure 3

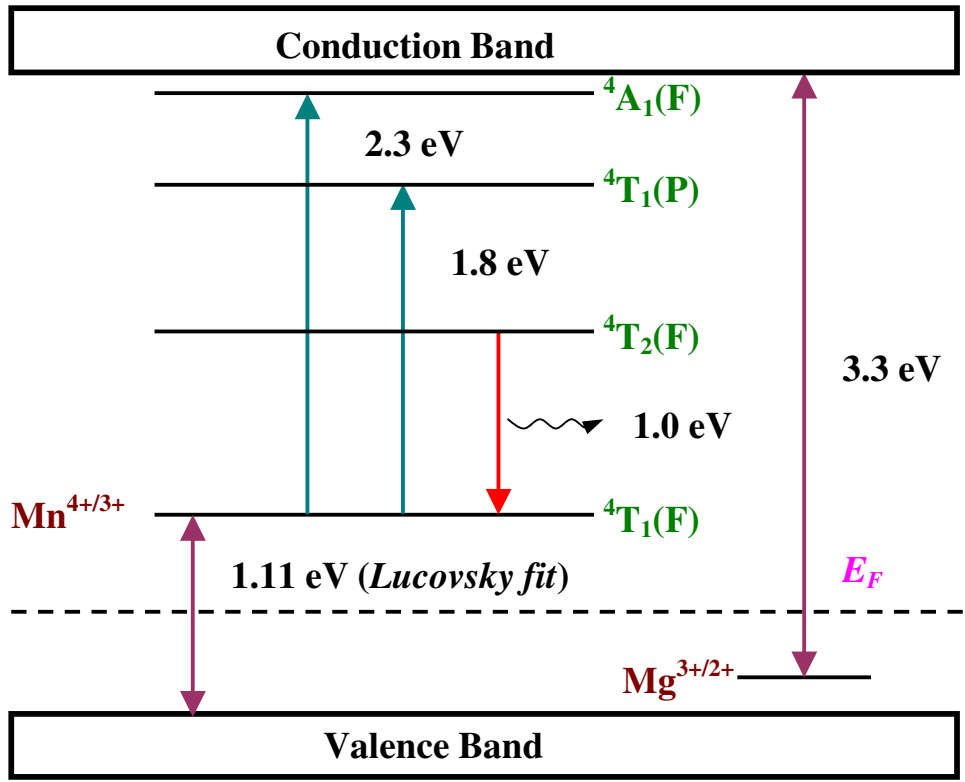